\documentclass[conference]{IEEEtran}
\usepackage{graphicx}
\usepackage{amsmath,amssymb,dsfont,bbm,epstopdf,pgfplots} 
\usepackage{amstext,amsfonts,amssymb,graphicx,stmaryrd}
\usepackage{color}
\usepackage{cite}
\usepackage{graphics}
\usepackage{tikz}
\usepackage{pgfplots}
\usepackage{stfloats}
\usetikzlibrary{shapes, arrows, decorations.markings, arrows.meta}
\usetikzlibrary{patterns} 
\allowdisplaybreaks
\graphicspath{{.}{./Figures/}} 
\allowdisplaybreaks[4] 
\usepackage{xpatch}
\usepackage{amscd}
\usepackage{epsfig}
\usepackage{graphics}
\usepackage{psfrag}
\usepackage{rotating}
\usepackage{amsmath}
\usepackage{amsfonts}
\usepackage{url}
\usepackage{color}
\usepackage[caption=false,font=footnotesize]{subfig}
\usepackage{epstopdf}
\usepackage{amsthm}
\usepackage{tikz}
\usepackage{mathtools}
\usetikzlibrary{shapes.geometric,arrows,shapes,decorations,backgrounds}
\usepackage[final]{pdfpages}
\newtheorem{theorem}{Theorem}
\newtheorem{lemma}[theorem]{Lemma}

\newtheorem{remark}{Remark}
\newtheorem{definition}{Definition}
\newtheorem{proposition}[theorem]{Proposition}

\newtheorem{corollary}[theorem]{Corollary}

\newfont{\bbb}{msbm10 scaled 500}

\newfont{\bb}{msbm10 scaled 1100}

\newcommand{\EE}{\mbox{\bb E}}

\newcommand{\onev}{{\bf 1}}

\newcommand{\CDc}{{\cal{CD}}}
\newcommand{\Cc}{{\cal C}}
\newcommand{\Dc}{{\cal D}}

\newcommand{\Sc}{{\cal S}}

\newcommand{\Wc}{{\cal W}}

\newcommand{\Yc}{{\cal Y}}
\newcommand{\Zc}{{\cal Z}}

\newcommand{\Pro}{{ \text{P}}}

\DeclareFontFamily{U}{cmfi}{}
\DeclareFontShape{U}{cmfi}{m}{n}{ <-> cmfi10 }{}
\DeclareSymbolFont{CMFI}{U}{cmfi}{m}{n}

\newcommand{\mkv}{-\!\!\!\!\minuso\!\!\!\!-}

\usepackage{multicol}
\begin{document}
	
	\title{Joint Sensing and Communication over Memoryless Broadcast Channels }
	
	
	\author{
		\IEEEauthorblockN{Mehrasa Ahmadipour\IEEEauthorrefmark{1}, Mich\`ele Wigger\IEEEauthorrefmark{1},  Mari Kobayashi\IEEEauthorrefmark{2} 
		}
		\IEEEauthorblockA{\small\IEEEauthorrefmark{1} LTCI Telecom Paris, IP Paris, 91120 Palaiseau, France, Emails:
			\url{{mehrasa.ahmadipour,michele.wigger}@telecom-paris.fr}}
		\IEEEauthorblockA{\small\IEEEauthorrefmark{2} Technical University of Munich, Munich, Germany,  Email: mari.kobayashi@tum.de
		}
	}
	
	\maketitle

	\begin{abstract} 
		A memoryless state-dependent broadcast channel (BC) is considered, where the transmitter wishes to convey two private messages to 
		two receivers while simultaneously estimating the respective states via generalized feedback. The model at hand is motivated by a joint radar and communication system where radar and data applications share the same frequency band. For physically degraded BCs with i.i.d. state sequences, we characterize the capacity-distortion tradeoff region. 
		For general BCs, we provide inner and outer bounds on the capacity-distortion region, as well as a sufficient condition when it is equal to the product of the capacity region and the set of achievable distortion. Interestingly, the proposed synergetic design significantly outperforms a conventional approach that splits the resource either for sensing or communication. 
	\end{abstract}
	
	\IEEEpeerreviewmaketitle

	\section{Introduction}
	A key-enabler of future high-mobility networks such as Vehicle-to-Everything (V2X) is the ability to continuously track the dynamically changing environment, hereafter called the \emph{state}, and to react accordingly by \emph{exchanging information} between nodes. 
	Although state sensing and communication have been designed separately in the past, power and spectral efficiency as well as 
	hardware costs encourage the integration of these two functions, such that they are operated by sharing the same frequency band and hardware (see e.g. \cite{zheng2019radar}). A typical example of such a scenario is joint radar parameter estimation and communication, where the transmitter equipped with a monostatic radar wishes to convey a message to a (already detected) receiver and simultaneously estimate the state parameters of interest such as velocity and range \cite{gaudio2019effectiveness}. Motivated by such an application, the first information theoretical model for joint sensing and communication has been introduced in \cite{kobayashi2018joint}. By modeling the backscattered signal as generalized feedback and designing carefully the input signal, the capacity-distortion tradeoff has been characterized for a single-user channel \cite{kobayashi2018joint}, while lower and upper bounds on the rate-distortion region over multiple access channel has been provided in \cite{kobayashi2019joint}. 

	The current paper extends \cite{kobayashi2018joint} to the broadcast channel (BC),  where the transmitter wishes to convey private messages to two receivers and simultaneously estimate their respective states. For simplicity, the state information is assumed known at each receiver. Although oversimplified, the scenario at hand relates to vehicular networks where a transmitter vehicle, equipped with a monostatic radar, sends (safety-related) messages to multiple vehicles and simultaneously estimates the parameters of these vehicles. The full characterization of the capacity-distortion region is very challenging, because the capacity region of memoryless BCs with generalized feedback is generally unknown even without state sensing (see e.g. \cite{shayevitz2012capacity}). Therefore, we consider first physically degraded BCs where generalized feedback is only useful for state sensing, like for the single user channel. The capacity-distortion region is completely characterized for this class of BCs. Moreover, closed-form expressions of the region are provided for some binary examples. The numerical evaluations illustrate interesting tradeoffs between the achievable rates and  distortions across two receivers. 
	For general BCs, we provide a sufficient condition when the capacity-distortion region is simply the product of the capacity region and the set of all achievable distortions, thus no tradeoff between communication and sensing arises. Furthermore, we provide general inner and outer bounds on the capacity-distortion region, as well as a state-dependent version of Dueck's BC. For all these kinds of BCs, 
	we show though numerical examples that the synergetic design significantly outperforms the resource-sharing scheme that splits the resource either for sensing or communication. 
	
	The rest of the paper is organized as follows. Section \ref{sect:Model} introduces our model and Section \ref{sec:notradeoff} presents some cases that yield no tradeoff between sensing and communication.  Section \ref{sect:DegradedBC} focuses on the physical degraded broadcast channel and provides some examples. Finally, upper and lower bounds for the general memoryless broadcast channel are provided along with an example in Section \ref{sect:GeneralBounds}. 

	\section{System Model}\label{sect:Model}
\begin{figure}[!t]
	\centering
	\includegraphics[scale=0.65]{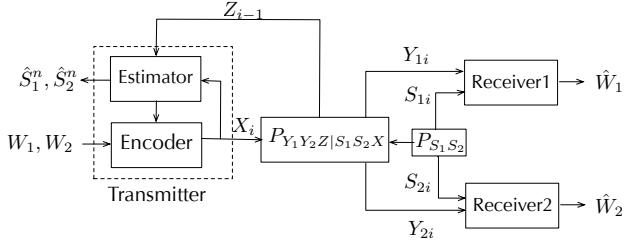}\caption{Broadcast model for joint sensing and communication}
		\label{fig:Model}
	\end{figure}

	\vspace{-0.2em}
	Consider a two-user state-dependent memoryless broadcast channel (SDMBC) with two private messages $W_1$ and $ W_2$ as illustrated in Fig.~\ref{fig:Model}. The model comprises a two-dimensional memoryless state sequence $\{(S_{1,i}, S_{2,i})\}_{i\geq 1}$  whose samples at time $i$ are distributed according to a given joint law $P_{S_1S_2}$ over the state alphabets $\mathcal{S}_1\times \mathcal{S}_2$. Given input and output alphabets $\mathcal{X}, \mathcal{Y}_1, \mathcal{Y}_2, \mathcal{Z}$,  input $X_i=x\in\mathcal{X}$ and state-realizations $S_{1,i}=s_1\in\mathcal{S}_1$ and $S_{2,i}=s_2\in\mathcal{S}_2$, the SDMBC produces a triple of outputs $(Y_{1,i}, Y_{2,i}, Z_i)\in \Yc_1 \times \Yc_2\times \Zc$ according to a given time-invariant transition law $P_{Y_1Y_2Z|S_1S_2X}(\cdot, \cdot, \cdot|s_1,s_2,x)$, for each time $i$. 
	A SDMBC is thus entirely specified by the tuple of alphabets and (conditional) pmfs 
	\begin{equation}(\mathcal{X}, \mathcal{Y}_1, \mathcal{Y}_2, \mathcal{Z}, P_{S_1S_2}, P_{Y_1Y_2Z|S_1S_2X}).
	\end{equation}
	We will often describe a SDMBC only by the pair of pmfs $(P_{S_1S_2}, P_{Y_1Y_2Z|S_1S_2X})$, in which case, the  corresponding alphabets should be clear from the context.
	
	A $(2^{nR_1},2^{nR_2}, n)$ code for an SDMBC  $P_{Y_1Y_2Z|S_1S_2X}$ consists of
	\begin{enumerate}
		\item two message sets $\Wc_1= [1:2^{nR_1}]$ and $\Wc_2= [1:2^{nR_2}]$;
		\item a sequence of encoding functions $\phi_i\colon \mathcal{W}_1\times \mathcal{W}_2 \times \mathcal{Z}^{i-1} \to \mathcal{X}$, for $i=1,2,\ldots,n$; 
		\item for each $k=1,2$ a decoding function  $g_k \colon \mathcal{S}_k^n \times \mathcal{Y}_k^n \to \mathcal{W}_k$; 
		\item for each $k=1,2$  a state estimator  $h_k \colon \mathcal{X}^n \times \mathcal{Z}^n \to \hat{\mathcal{S}}_k^n$, where  $\hat{\mathcal{S}}_k$ denotes the given  reconstruction alphabet for state sequence $S^n_k=(S_{k,1},\cdots,S_{k,n})$.
	\end{enumerate}
	
	For a given code, we let  the random messages $W_1$ and $W_2$ be uniform over the message sets $\mathcal{W}_1$ and $\mathcal{W}_2$ and the inputs $X_i=\phi_i(W_1,W_2, Z^{i-1})$, for $i=1,\ldots, n$. The corresponding outputs $Y_{1,i} Y_{2,i}, Z_i$ at time $i$ are obtained from the states $S_{1,i}$ and $S_{2,i}$  and the input $X_i$ according to the  SDMBC transition law $P_{Y_1Y_2Z|S_1S_2X}$. Further, let $\hat{S}_k^n:=(\hat{S}_{k,1},\cdots,\hat{S}_{k,n} )=h_k(X^n, Z^n)$ be the state estimates at the transmitter and let $\hat{W}_k=g_k(S_k^n,Y_k^n)$ be the decoded message by decoder $k$, for $k=1,2$.
	
	The quality of the state estimates $\hat{S}_k^n$  is measured by  a given per-symbol distortion function $d_k\colon \Sc_k\times \hat{\Sc}_k \mapsto [0,\infty)$, and we will be interested in the \emph{expected average per-block distortion}
	\begin{equation}
	\Delta_k^{(n)}\triangleq \frac{1}{n} \sum_{i=1}^n \EE[d_k(S_{k,i}, \hat{S}_{k,i})], \quad k=1,2.
	\end{equation}
	For the decoded messages $\hat{W}_1$ and $\hat{W}_k$ we focus  on their joint probability of error:
	\begin{equation}
	p^n(\textnormal{error}) := \textnormal{Pr}\left(\hat{W}_1\neq W_1 \quad \textnormal{or} \quad \hat{W}_2 \neq W_2\right).
	\end{equation}
	\begin{definition} 
		A rate-distortion tuple $(R_1, R_2, D_1, D_2)$ is said   achievable if there exists  a sequence (in $n$) of  $(2^{nR_1},2^{nR_2}, n)$ codes that simultaneously satisfy
		\begin{subequations}\label{eq:asymptotics}
			\begin{IEEEeqnarray}{rCl}
				\lim_{n\to \infty}	p^{(n)}(\textnormal{error}) &=&0 \\
				\varlimsup_{n\to \infty}	\Delta_k^{(n)}& \leq& D_k, \quad \textnormal{for } k=1,2.
			\end{IEEEeqnarray}
		\end{subequations}
	\end{definition}
	The closure of the union of all achievable rate-distortion tuples $(R_1, R_2,D_1,D_2)$ is called the capacity-distortion region and is denoted $\CDc$. 
	The current work aims at specifying the tradeoff between the achievable rates and distortions. 
	As we will see in Sections ~\ref{sec:notradeoff} and \ref{sect:GeneralBounds}, there is no such tradeoff in some cases, and the resulting region $\CDc$ is the product of  SDMBC's capacity region:
	\begin{IEEEeqnarray}{rCl}
		\mathcal{C}\triangleq \{(R_1, R_2)\colon (R_1,R_2,D_1,D_2) \in \CDc \; \textnormal{for } D_1,D_2 \geq 0\},\IEEEeqnarraynumspace
	\end{IEEEeqnarray}
	and its distortion region:
	\begin{IEEEeqnarray}{rCl}
		\mathcal{D}\triangleq \{(D_1, D_2)\colon (R_1,R_2,D_1,D_2) \in \CDc \; \textnormal{for } R_1,R_2 \geq 0\}.\IEEEeqnarraynumspace
	\end{IEEEeqnarray}
	
	Before presenting our results on the tradeoff region $\CDc$ in the following sections, we describe the optimal choice of the estimators $h_1$ and $h_2$.
	\begin{lemma}\label{lemma:Shat} 
		For  $k=1,2$ and any $i=1,\ldots, n$, whenever $X_i=x$ and $Z_i=z$, the optimal estimator $h_k$ that minimizes the average expected distortion $\Delta_k^{(n)}$
		is given by 
		\begin{IEEEeqnarray}{rCl}
			\hat{s}_{k,i}^*(x,z) &\triangleq & {\rm arg}\min_{s'\in \hat{\Sc}_k} \sum_{s_k\in \Sc_k} P_{S_{k,i}|X_iZ_i}(s_k|x,z)  d(s_k, s'). \IEEEeqnarraynumspace
			\label{eq:estimator}
		\end{IEEEeqnarray}
		In above definition \eqref{eq:estimator}, ties can be broken arbitrarily.
	\end{lemma}
	Notice that the lemma implies in particular that a symbolwise estimator that estimates $S_{k,i}$ only based on $(X_i,Z_i)$ is  optimal; there is no need to resort to previous or past observations $(X^{i-1},Z^{i-1})$ or $(X_{i+1}^n,Z_{i+1}^n)$.
	\renewcommand{\qedsymbol}{}
	\begin{IEEEproof}[Proof of Lemma~\ref{lemma:Shat}] Recall that $\hat{S}_k^n$ is a function of $X^n, Z^n$ and write for each $i=1,\cdots,n$:
		\begin{IEEEeqnarray}{rCl}\label{eq:step1}
			\lefteqn{\EE\left[d_k(S_{k,i}, \hat{S}_{k,i})\right] } \quad \nonumber \\
			&=& \EE_{X^n,Z^n}\left[ \EE[d_k(S_{k,i}, \hat{S}_{k,i}) |X^n,Z^n] \right]  
			\\
			&\overset{\mathrm{(a)}}=&\sum_{x^n,z^n} P_{X^nZ^n}(x^n,z^n) \sum_{\hat{s}_k\in \Sc_k}  P_{\hat{S}_{k,i}|X^nZ^n} (\hat{s}_k|x^n,z^n)
			\nonumber\\
			&&\hspace{1cm} \cdot \sum_{s_k} P_{S_{k,i}|X_iZ_i}(s_k|x_i,z_i)  d(s_k, \hat{s}_k)
			\\
			&\geq& \sum_{x^n,z^n} P_{X^nZ^n}(x^n,z^n) 		\nonumber\\
			&&\hspace{1cm} \cdot \min_{\hat{s}_k\in \Sc_k}\sum_{s_k} P_{S_{k,i}|X_iZ_i} (s_k|x_i, z_i)  d(s_k, \hat{s}_k) 
			\nonumber\\
			&=& \EE[d(S_{k,i}, \hat{s}^*_{k,i}(X_i,Z_i))],
		\end{IEEEeqnarray}
		where $(a)$ holds by the Markov chain 
		\begin{equation*}
		\Big(X^{i-1},X_{i+1}^{n}, Z^{i-1},Z_{i+1}^n, \hat{S}_{k,i}\Big) \mkv (X_i,Z_i)\mkv S_{k,i}.
		\end{equation*}
	\end{IEEEproof}
	\section{Absence of Rate-Distortion Tradeoff}\label{sec:notradeoff}
	We first consider degenerate cases where the rate-distortion region $\CDc$ is given by the Cartesian product between the capacity region $\Cc$ and the distortions region $\Dc$.  
	\begin{proposition}[No Rate-Distortion Tradeoff]\label{prp1}
		Consider a SDMBC $( P_{S_1S_2}, P_{Y_1Y_2Z|S_1S_2X})$ and let $(X,S_1,S_2,Y_1,Y_2,Z) \sim P_{X} P_{S_1S_2} P_{Y_1Y_2Z|S_1S_2X}$ for a given input law $P_X$. If there exist functions $\psi_1$ and $\psi_2$ with domain $\mathcal{Z}$ such that for all $P_X$  the Markov chains 
		\begin{IEEEeqnarray}{rCl}
			&(S_k,\psi_{k}(Z)) \perp  X,\label{cond1}\\
			&S_k\mkv \psi_{k}(Z)\mkv (Z,X), \quad k\in\{1,2\}, \label{cond2}
		\end{IEEEeqnarray} 
		hold, then for the SDMBC under consideration:
		\begin{IEEEeqnarray}{rCl}\label{eq:product}
			\CDc =  \mathcal{C} \times \mathcal{D}.
		\end{IEEEeqnarray}	
		In this case, there is no tradeoff between the achievable rate pairs $(R_1, R_2)$ and the achievable distortion pairs $(D_1, D_2)$.
	\end{proposition}
	\begin{IEEEproof}
		Notice under the given Markov chain \eqref{cond2}:
		\begin{equation}\label{eq:d}
		P_{S_{k,i}|X_iZ_i}(s_k|x,z)=P_{S_{k,i}|\psi_{k}(Z_i)}(s_k|\psi_k(z)). 
		\end{equation}
		Trivially, $\CDc \subseteq  \mathcal{C} \times \mathcal{D}$. To see that also $\CDc \supseteq  \mathcal{C} \times \mathcal{D}$ holds, notice that by \eqref{eq:d} and Lemma~\ref{lemma:Shat} the optimal estimators depend only on the sequences $\{\psi_k(Z_i)\}_{i=1}^n$, for $k=1,2$, and thus by \eqref{cond1} are independent of the considered coding scheme and the produced inputs. 
	\end{IEEEproof}
	In the following corollary, 
	The following example satisfies conditions \eqref{cond1} and \eqref{cond2} in Proposition~\ref{prp1} for an appropriate choice of $\psi_1$ and $\psi_2$.
	\subsection{Example: Erasure BC with Noisy Feedback}
	Let the joint  law $\Pro_{S_1S_2E_1E_2}(s_1, s_2, e_1, e_2)$ over $\{0,1\}^4$ be arbitrary but given, and $(E_1,E_2, S_1, S_2) \sim  \Pro_{S_1S_2E_1E_2}$. 
	Consider the  state-dependent erasure BC
	\begin{IEEEeqnarray}{rCl}
		Y_{k}=\begin{cases}
			&X \quad \text{if }  S_{k}=0,\\
			&? \quad \text{if } S_{k}=1,
		\end{cases}\qquad {k\in\{1,2\}},
	\end{IEEEeqnarray}
	where the feedback signal $Z=(Z_1,Z_2)$ is given by
	\begin{IEEEeqnarray}{rCl}
		Z_k=\begin{cases}
			&Y_{k} \quad \text{if }  E_k=0,\\
			&? \quad \text{if } E_k=1,
		\end{cases}\qquad {k\in\{1,2\}}.
	\end{IEEEeqnarray}
	Further consider  the Hamming distortion measure $d_k(s, \hat{s}) = s \oplus \hat{s}$, for $k=1,2$. For the choice
	\begin{equation}
	\psi_k(Z)=
	\begin{cases}
	&1, \qquad \text{if} \quad Z_k=?\\
	&0,  \qquad \text{else}, 
	\end{cases}
	\end{equation}
	the described SDMBC satisfies the conditions in Proposition~\ref{prp1} and its capacity-distortion region is thus given by
	\begin{equation}
	\CDc = \Cc \times \Dc.
	\end{equation}
	\begin{remark}
		For the case of output feedback $Z=(Y_1, Y_2)$ or $E_1=E_2=0$, the transmitter can perfectly estimate the state $(S_1, S_2)$, yielding $D_1=D_2=0$ regardless of the rate pair $(R_1, R_2)\in \Cc$. 
		The capacity region $\Cc$ of the erasure broadcast channel with output feedback is still unknown in general.
	\end{remark}
	
	\section{Physically Degraded BCs}\label{sect:DegradedBC}
	In this section, by focusing on the physically degraded SDMBC, we fully characterize the capacity-distortion region. Then, we discuss two binary physically degraded SDMBCs to illustrate the rate-distortion tradeoff between the two receivers. 
	\begin{definition}
		An SDMBC $( P_{S_1S_2},$ $P_{Y_1Y_2Z|S_1S_2X})$  is called \emph{physically degraded} if  there are conditional laws $P_{Y_	1|XS_1}$ and $P_{Y_2S_2|  S_1 Y_1}$ such that 
		\begin{equation}
		P_{Y_1Y_2|S_1S_2X} P_{S_1S_2} = P_{S_1}P_{Y_1|S_1X} P_{Y_2S_2|  S_1 Y_1}.
		\end{equation}
		That means for any arbitrary input $P_X$, if  a tuple $(X,S_1,S_2,Y_1,Y_2)\sim P_{X} P_{S_1S_2} P_{Y_1Y_2|S_1S_2X}$, then it satisfies the Markov chain 
		\begin{equation}\label{eq:Mc}
		X \mkv (S_1, Y_1) \mkv (S_2, Y_2). 
		\end{equation}
	\end{definition}
	\begin{proposition}\label{propose1}
		The  capacity-distortion region $\Cc\Dc$ of a physically degraded SDMBC is the closure of the set of all quadruples $(R_1,R_2,D_1,D_2)$ for which there exists a joint law $P_{UX}$ so that the tuple $(U, X, S_1, S_2, Y_1, Y_2, Z)\sim P_{UX} P_{S_1S_2} P_{Y_1Y_2Z|S_1S_2X}$ satisfies  the two rate constraints 
		\begin{align}
		R_1&\leq I(U;Y_1|S_1)\label{R11}\\
		R_2&\leq I(X;Y_2\mid S_2, U),\label{R22}
		\end{align}
		and the distortion constraints
		\begin{equation}
		\mathbb{E}[d_k(S_k, \hat{s}_{k}^*(X,Z)))]\leq D_k, \quad k\in\{1,2\},\\
		\end{equation}
		where 
		\begin{IEEEeqnarray}{rCl}
			\hat{s}_{k}^*(x,z) &\triangleq & {\rm arg}\min_{s'\in \hat{\Sc}_k} \sum_{s_k\in \Sc_k} P_{S_{k}|XZ}(s_k|x,z)  d(s_k, s'). \IEEEeqnarraynumspace
			\label{eq:est}
		\end{IEEEeqnarray}
		
	\end{proposition}
	\begin{IEEEproof} The converse follows as a special case of Theorem~\ref{outer1} ahead where one can ignore constraints \eqref{upper3} and \eqref{upper4}. Notice that constraint \eqref{R2} is equivalent to \eqref{R22} because $(U,X)$ is independent of $(S_1,S_2)$ and because for a physically degraded DMBC the Markov chain \eqref{eq:Mc} holds.  
		
		Achievability is obtained by simple superposition coding and using the optimal estimator described in Lemma~\ref{lemma:Shat}.
	\end{IEEEproof}
	
	We consider two binary state-dependent channels. For the binary states, we consider the Hamming distortion measure.
	\subsection{Example: Binary BC with Multiplicative States}\label{ex:Binary1}
	Consider the physically degraded SDMBC with binary input/output alphabets $\mathcal{X}=\mathcal{Y}_1=\mathcal{Y}_2=\{0,1\}$ and binary
	state alphabets $\mathcal{S}_1=\mathcal{S}_2=\{0,1\}$. The channel input-output relation is described by 
	\begin{IEEEeqnarray}{rCl}\label{ex2}
		Y_k&=& X \cdot S_k, \qquad k=1,2 , 
	\end{IEEEeqnarray}
	with 
	the joint state pmf 
	\begin{align}\label{ew:p_ss}
	P_{S_1 S_2} (s_1,s_2) &= \begin{cases}
	1-q, & \text{if $(s_1, s_2)= (0,0)$} \\
	0, & \text{if $(s_1, s_2)= (0,1)$}\\
	q\cdot \gamma, & \text{if $(s_1, s_2)= (1, 1)$} \\
	q\cdot (1-\gamma)& \text{if $(s_1, s_2)= (1, 0)$},
	\end{cases}
	\end{align}
	for 
	$\gamma, q\in[0,1]$.
	Notice that $S_2$ is a degraded version of $S_1$.
	We consider output feedback $Z=(Y_1, Y_2)$. 
	\begin{corollary}\label{prop:binary1}				
		The capacity-distortions region $\CDc$ of the binary physically degraded SDMBC in 
		\eqref{ex2}--\eqref{ew:p_ss} is  the set of all quadruples $(R_1,R_2,D_1,D_2)$ satisfying 
		\begin{subequations}\label{eq:con}
			\begin{IEEEeqnarray}{rCl}
				R_1&\leq&
				q\cdot H_\textnormal{b}(p)\cdot r\label{R1f}\\
				R_2&\leq &
				\gamma \cdot q\cdot H_\textnormal{b}(p)\cdot (1-r)\label{R2f}\\
				D_1 &\geq & (1-{p})\cdot \min\{q, 1-q\} \label{eq:conD1}\\
				D_2 &\geq &(1-{p}) \cdot \min\{\gamma\cdot q, 1-\gamma \cdot q \} , \label{eq:conD2}
			\end{IEEEeqnarray}
		\end{subequations}
		for some choice of the parameters $r, p\in[0,1]$. 
	\end{corollary}
	
	\begin{proof}
		It suffices to evaluate the rate-constraints \eqref{R11} and \eqref{R22} for  $X=V\oplus U$ when $U$ and $V$ are 
		independent Bernoulli distributed random variables. In \eqref{eq:con}, we choose the parameter $p=\Pr[X=1]$ and $r=1-\frac{H(V)}{H_b(p)}$.
		To calculate the distortion, we determine the optimal estimator $\hat{s}^*_{k}(x,y_1, y_2)$ from \eqref{eq:est} as
		\begin{subequations}
			\begin{IEEEeqnarray}{rCl}\label{eq:estimator1}
				\hat{s}^*_k(1, y_1, y_2) &=& y_k, \\
				\hspace{1cm}  \hat{s}^*_k(0, y_1, y_2) &=& \onev\{ P_{S_k}(1) >1/2 \}.  \hspace{1cm} \hfill\qedhere
			\end{IEEEeqnarray}
		\end{subequations}
	\end{proof}	
	\begin{remark}
		Fixing $r=1$, the capacity-distortion region in \eqref{eq:con} reduces to the capacity-distortion tradeoff of a single user channel \cite[Proposition 1]{kobayashi2018joint}. Similarly to the single-user case, we observe the tension between the minimum distortion by choosing $p=1$ (always sending $X=1$) and the maximum rate by choosing $p=1/2$. In the BC, the resource is shared between the two users via the time-sharing parameter $r$. 
	\end{remark}	
		\begin{figure}[!t]
		\vspace{-0.3cm}
		\centering
		\includegraphics[width=0.5\textwidth]{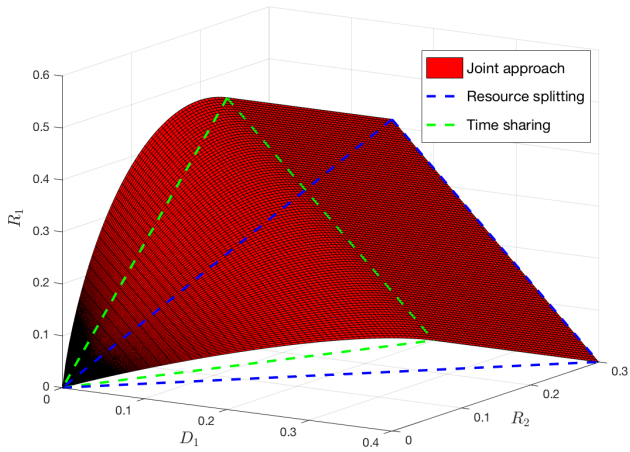}
		\vspace{-0.5cm}
		\caption{Boundary of the capacity-distortion region $\Cc\Dc$ for the  example in Subsection~\ref{ex:Binary1}.}
		\vspace{-0.5cm}
		\label{fig:binary}
	\end{figure}
	
We evaluate the capacity-distortion region \eqref{eq:con} for $\gamma=0.5$ and $q=0.6$.
	Fig.~\ref{fig:binary} shows in red colour the dominant boundary points of the projection of the tradeoff region $\CDc$ onto the $3$-dimensional plane $(R_1, R_2, D_1)$. 
	 The tradeoff with $D_2$ is omitted because $D_2$ is a scaled version of $D_1$. 	
	 	 
		It is worth comparing the capacity-distortion region $\CDc$, achieved by the proposed \emph{co-design scheme}
	that uses a common waveform for both sensing and communication tasks,  with the rate-distortion region achieved by a baseline scheme, called \emph{resource splitting}, that  separates the two tasks into two modes. In  the \emph{sensing mode}, the transmitter estimates the states via the feedback but does not communicate any messages to the two receivers. In the \emph{communication mode}, it communicates with the receivers but without using the feedback. Moreover, in this second mode,  it also estimates the states but again without accessing the feedback. 
	
	For the example at hand,  the resource splitting scheme acts as follows. During the sensing mode, the transmitter always sends $X=1$ (which is equivalent to setting $p=1$ in \eqref{eq:con}) so as to minimize the distortion. This achieves 
	\begin{equation}\label{eq:t1}
	(R_1, R_2, D_1, D_2)= (0, 0, 0,0).
	\end{equation}
	During the communication mode, the transmitter sets $p=0.5$ in \eqref{eq:con}\footnote{Recall that the capacity-distortion region in \eqref{eq:con} is achieved without using the feedback for communication because the BC is physically degraded.} so as to maximize the communication rate and without using the feedback it estimates the states as 
	$\hat{s}_1=\onev\{q>0.5\}$ and $\hat{s}_2= \onev\{q\cdot\gamma >0.5\}$.
	 This  achieves 
	 \begin{equation}\label{eq:t2}
	 (R_1, R_2, D_1 ,D_2)=(q\cdot r, \gamma\cdot q\cdot(1-r), D_{1, \max}, D_{2, \max})
	 \end{equation} where  $D_{1, \max}=\min\{q, 1-q\}$ and $D_{2, \max}= \min\{\gamma\cdot q, 1-\gamma\cdot q\}$, and  where $r\in [0,1]$ denotes the time-sharing parameter between the two two communication rates. Fig.~\ref{fig:binary} shows the time-sharing region between the two modes \eqref{eq:t1} and \eqref{eq:t2} in blue colour. 
	 
	 Fig.~\ref{fig:binary} also shows the region achieved by a more sophisticated time-sharing scheme that combines  the minimum distortion point 
	  $(R_1, R_2, D_1, D_2)=(0, 0, 0, 0)$ of the capacity-distortion region $\CDc$ with  the maximum communication rate points  of $\CDc$, $(R_1, R_2, D_1 ,D_2)=(q\cdot$$ r, \gamma\cdot q \cdot (1-r), \frac{D_{1, \max}}{2}, \frac{D_{2, \max}}{2})$ for $r\in[0,1]$.
	
	We observe that both resource splitting and time sharing approaches fail to achieve the entire region $\CDc$.

	So far, there was no tradeoff between the two distortion constraints $D_1$ and $D_2$. This is different in the next example, which otherwise is very similar. 	
	\subsection{Example: Binary BC with Flipping Inputs}\label{ex:BinaryFlipped}
	\newcommand{\bp}{{\bar{p}}}
	\newcommand{\bq}{{\bar{p_s}}}
	\newcommand{\bg}{{\bar{\gamma}}}
	Reconsider the same state pmf $P_{S_1S_2}$ as in the previous example, but  now a SDMBC with transition law 
	\begin{IEEEeqnarray}{rCl}\label{model:flipped}
		Y_1&=& X \cdot S_1, \qquad Y_2 = (1-X)\cdot S_2.
	\end{IEEEeqnarray}
	Consider output feedback $Z=(Y_1,Y_2)$. 
	\begin{corollary}\label{prop:binary2}		
		The capacity-distortion region $\CDc$ of the binary SDMBC with flipping inputs in \eqref{model:flipped} and output feedback is the set of all quadruples $(R_1,R_2,D_1,D_2)$ satisfying				\begin{subequations}\label{eq:con2}
			\begin{IEEEeqnarray}{rCl}
				R_1 &\leq& q\cdot H_\textnormal{b}(p)\cdot r\label{r1} \\
				R_2 &\leq&\gamma \cdot q\cdot H_\textnormal{b}(p)\cdot (1-r) \label{r2} \\
				D_1&\geq&(1-p) \cdot  \min \{ q (1-\gamma),(1-q)\} \label{d1}\\
				D_2&\geq&p \cdot q \min \{  \gamma, 1-\gamma \}\label{d2}
			\end{IEEEeqnarray}
		\end{subequations}
		for some choice of the parameters $r, p\in[0,1]$. 
	\end{corollary}				
	\begin{proof}
		To achieve this region, we can consider the same choices of $(U,X)$ as in the previous example.  
		The optimal estimators are given by \eqref{eq:estimator1} for receiver 1 and 
		\begin{subequations} 
		\begin{IEEEeqnarray}{rCl}\label{eq:estimator2}
			\hat{s}^*_2(0, y_1, y_2) &=& y_2, \\
			\hat{s}_2^*(1, y_1, y_2) &=& \onev\{ P_{S_2}(1) >1/2 \}\qedhere
		\end{IEEEeqnarray}
		\end{subequations}
		for receiver 2. Contrary to the previous example, we observe a tradeoff between the achievable distortions $D_1$ and $D_2$. 
		%
	\end{proof}		

	\section{General BCs}\label{sect:GeneralBounds}
		\begin{figure*}[ht]
			
			\begin{subequations}\label{eq:inner}
				\begin{IEEEeqnarray}{rCl}
					R_1 &\leq& I(U_0,U_1;Y_1,V_1\mid S_1) - I(U_0,U_1,U_2, Z;V_0,V_1|S_1,Y_1)\label{eq:inner1}
					\\
					R_2 &\leq& I(U_0,U_2; Y_2,V_2\mid S_2) - I(U_0,U_1,U_2, Z;V_0,V_2|S_2,Y_2)\label{eq:inner2}
					\\
					R_1+R_2 &\leq& I(U_1; Y_1,V_1|U_0, S_1) + I(U_2; Y_2,V_2|U_0, S_2) + \min_{i\in\{1,2\}}I(U_0;Y_i,V_i\mid S_i)- I(U_1;U_2|U_0)\nonumber \\
					&&\hspace{-1cm}-I(U_0,U_1,U_2, Z;V_1|V_0,S_1,Y_1) -I(U_0,U_1,U_2, Z;V_2|V_0,S_2,Y_2) - \max_{i\in\{1,2\}} I(U_0,U_1,U_2, Z;V_0|S_i,Y_i)
				\end{IEEEeqnarray}
			\end{subequations}
			\hrule
		\end{figure*}
\subsection{General Bounds}
	Reconsider  the general SDMBC (not necessarily physically degraded). We provide an inner and an outer bound on the capacity-distortion region.
	\begin{theorem}\label{outer1}
		If $(R_1, R_2, D_1, D_2)$ is achievable on a SDMBC $(P_{S_1S_2},P_{Y_1Y_2Z|S_1S_2X})$, then there exists  for each $k=1,2$  a conditional pmf $P_{U_k|X}$ such that the random tuple $(U_k, X, S_1,S_2,Y_1,Y_2, Z)\sim P_{U_k|X}P_X P_{S_1S_2} P_{Y_1Y_2Z\mid S_1S_2X}$ satisfies the  rate constraints 
			\begin{subequations}\label{eq:Rupper}
			\begin{IEEEeqnarray}{rCl}
				R_1 &\leq &I(U_1;Y_1 \mid S_1),\label{R1}\\
				R_1+R_2 &\leq &I(X;Y_1, Y_2 \mid  S_1, S_2 , U_1),\label{R2}\\
				R_1+R_2 &\leq &I(X;Y_1, Y_2 \mid S_1, S_2, U_2), \label{upper3}\\
				R_2&\leq &I(U_2;Y_2 \mid S_2)\label{upper4}
			\end{IEEEeqnarray}
		\end{subequations}
		and the average distortion constraints
		\begin{equation}
		\mathbb{E}[d_k(S_k, \hat{s}_{k}^*(X,Z)))]\leq D_k, \quad k\in\{1,2\},\\
		\end{equation}
		where the function $\hat{s}_{k}^*(\cdot, \cdot)$ is defined in \eqref{eq:est}.
	\end{theorem}
	\begin{IEEEproof}  {See Appendix~\ref{app:converse_proof}.}
	\end{IEEEproof}

Achievability results are easily obtained by combining existing achievability results for SDMBCs with generalized feedback with the optimal estimator in Lemma~\ref{lemma:Shat}. For example,  
based on  \cite{shayevitz2012capacity} we obtain: 
	\begin{proposition}\label{prp:inner}
		Consider a SDMBC   $(P_{S_1S_2},P_{Y_1Y_2Z|S_1S_2X})$.
		\\ For any (conditional) pmfs $P_{U_0U_1U_2X}$ and $P_{V_0V_1V_2|U_0U_1U_2Z}$ and tuple
		$(U_0,  U_1,U_2,X,S_1,S_2, Y_1,Y_2,Z, V_0,V_1,V_2)\sim$ 
		$ P_{U_0U_1U_2X} P_{S_1S_2}P_{Y_1Y_2Z|S_1S_2X}   P_{V_0V_1V_2|U_0U_1U_2Z}$, 
		the {convex} closure  of the set of all quadruples $(R_1,R_2,D_1,D_2)$ satisfying   inequalities \eqref{eq:inner} on top of the this page and  the distortion constraints 
		\begin{equation}
		\mathbb{E}[d_k(S_k, \hat{s}_{k}^*(X,Z)))]\leq D_k, \quad k\in\{1,2\},\\
		\end{equation}
		for  $\hat{s}_{k}^*(\cdot, \cdot)$  defined in \eqref{eq:est}, is achievable.
			\end{proposition}

	\subsection{Example: Dueck's BC with Binary States}

\thicklines

\tikzset{XOR/.style={draw,very thick, circle,append after command={

        [very thick,shorten >=\pgflinewidth, shorten <=\pgflinewidth,]
        (\tikzlastnode.north) edge (\tikzlastnode.south)
        (\tikzlastnode.east) edge (\tikzlastnode.west)

        }
    }
}


\tikzstyle{startstop} = [rectangle, rounded corners, minimum width=3cm, minimum height=.1cm,text centered, draw=black, fill=red!30]

\begin{figure}[h!]
	\begin{center}
\begin{tikzpicture}[scale=0.35,node distance=2cm]
\node at (0,0)(input1) { $X_1$};
 \node   at (3,0) (XOR-aa)[XOR,scale=1.7] {};
 \node at (3,3)  (N) {$N$};
  \draw[line width=0.5mm] [->] (N)--(XOR-aa);
  
\draw[line width=0.5mm] (7,0) circle (.8cm);
 \node  at (7,0) [circle,scale=1.2](dot) {\textbullet};
 \node at (7,3)(s1) {
$S_1$};
  \draw[line width=0.5mm] [->] (s1)--(dot);

\node at (10,2)(y1) {
$Y_1'$};
 \draw[line width=0.5mm] (input1) -- (XOR-aa);
 \draw[line width=0.5mm] (XOR-aa)--(dot);
  \draw[line width=0.5mm][->](dot) -- (y1);

\node at (0,-2)(input0){
$X_0$};
\node at (10,0)(y01) {
$Y_0$};
\node at (10,-4)(y02) {
$Y_0$};
\draw[line width=0.5mm]  (input0) -- (7,-2);
\draw[line width=0.5mm] [->] (7,-2) -- (y01);
\draw[line width=0.5mm] [->] (7,-2) -- (y02);

\node at (0,-4)(input2){
$X_2$};

 \node  at (3,-4) (XOR-a)[XOR,scale=1.7] {};
 \node  at (3,-7) (N2) {$N$};
 \draw[line width=0.5mm] [->] (N2)--(XOR-a);
\node at (10,-6)(y2) {
$Y_2'$};

\draw[line width=0.5mm] (7,-4)  circle (0.8cm);
\node  at (7,-4) [circle,scale=1.2] (dott){\textbullet};
\node  at (7,-7) (S2) {
$S_2$};

\draw[line width=0.5mm] [->] (S2)--(dott);
  \draw[line width=0.5mm] (input2) -- (XOR-a);
       \draw[line width=0.5mm] (XOR-a)--(dott);
           \draw[line width=0.5mm][->] (dott) -- (y2);

\draw[line width=0.5mm,dashed] (-1,-5) rectangle(1,1);
\node at (-4,-2){
Transmitter};
\draw[line width=0.5mm,dashed] (9,-1) rectangle(11,3);
\node at (14,1){
Receiver1};
\draw[line width=0.5mm,dashed] (9,-7) rectangle(11,-3);
\node at (14,-5){
Receiver2};
\end{tikzpicture}
\end{center}

	\caption{A state-dependent version of Dueck's BC.}
	\label{fig:Dueck}
	\end{figure}
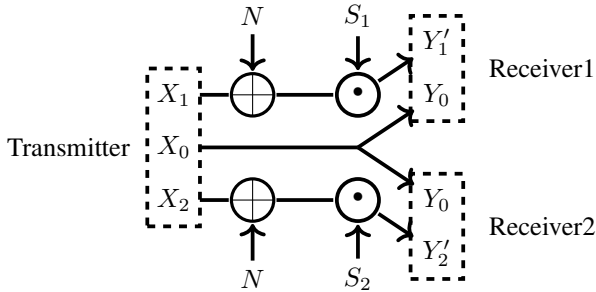
Consider  the state-dependent version of  Dueck's BC  \cite{DUECK} in Figure~\ref{fig:Dueck} 
	with input  $X=(X_0,X_1,X_2)\in \{0,1\}^3$ and  outputs 
	\begin{equation}
	Y_k=(X_0, Y_k', S_1,S_2), \qquad  {k\in\{1,2\}},
	\end{equation}
	for states $S_1,S_2 \in\{0,1\}$, 
	\begin{subequations}\label{eq:Mehrasa}
		\begin{equation}
		Y_k' =  S_k (X_k \oplus N), \qquad {k\in\{1,2\}},
		\end{equation}
	\end{subequations}
	and  $N$ a Bernoulli-$\frac{1}{2}$ noise independent of the inputs and the states. Assume i.i.d. states such that $P_{S_1S_2}(s_1,s_2)= P_{S}(s_1)P_{S}(s_2)$ for a given pmf $P_{S}$. 
	The feedback signal is  
	\begin{equation}
	Z= (Y_1', Y_2').
	\end{equation}
	
	Notice that in this example, only the input bits $X_1$ and $X_2$ are corrupted by the state and the noise, but not  $X_0$. This latter is thus completely useless for sensing. In fact, as we will show, for sensing it is optimal to  choose  $X_0$ arbitrary and depending on the state distribution either $X_1=X_2$ or $X_1\neq X_2$.  In contrast, for communication without feedback, it is optimal to send uncoded bits using $X_0$ and to disregard the other two input bits $X_1$ and $X_2$. The baseline resource splitting  scheme (where feedback is only used for sensing) thus orthogonalizes the inputs: $X_0$ is used for communication and $X_1,X_2$ are used for sensing. In a  traditional resource splitting scheme, the two modes  are never combined, which for this example is clearly suboptimal  because both modes (sensing and communication) can be  performed simultaneously without disturbing each other. As we will see, in certain cases (depending on the state distribution $P_S$) the simple   approach that performs both resource splitting modes simultaneously is optimal when one insists on achieving the smallest possible distortions. For larger distortions, it can however be improved by also exploiting the feedback and the inputs $X_1$ and $X_2$ for communication. This is for example achieved by the scheme leading to Propostion~\ref{prp:inner}, as we show in the following corollary and the subsequent numerical evaluation.

	\begin{corollary}\label{ex:general}
		The capacity-distortion region $\CDc$ of Dueck's state-dependent BC  is included in the set of  quadruples $(R_1,R_2,D_1,D_2)$ that for some choice of the parameters $p, q, \beta \in [0,1]$ satisfy the rate-constraints 	
		\begin{subequations}\label{outer}
			\begin{IEEEeqnarray}{rCl}
				R_1&\leq& 1-p\\
				R_2&\leq& p+(P_{S}(1))^2\cdot H_\textnormal{b}(\beta)\\
				R_1&\leq& q+(P_{S}(1))^2\cdot H_\textnormal{b}(\beta) \label{eq:R1-outer}\\ 
				R_2&\leq& 1-q\label{eq:R2-outer}
			\end{IEEEeqnarray}
			and  for each $k\in\{1,2\}$ the distortion constraint
			\begin{IEEEeqnarray}{rCl}\label{ex:distortin}
				D_k & \geq & \frac{1}{2}(1-\beta)\cdot \min\{P_S(1), P_{S}(0)\cdot(1+P_{S}(0))\} \nonumber
				\\
				&&+\frac{1}{2}\beta P_{S}(1)  \big[P_{S}(0)+ \min\{ P_S(0), P_S(1)\} \big]. \IEEEeqnarraynumspace\label{eq:Dh}
			\end{IEEEeqnarray}
		\end{subequations}
Moreover, depending on the values of $P_S(0)$ and $P_S(1)$, the following holds:
\begin{itemize}
\item When $P_S(1) \leq P_S(0)$, distortion constraint \eqref{ex:distortin} simplifies to $D_k\geq \frac{1}{2} P_S(1)= D_{\min}$ and one can restrict to $\beta=1$ in above outer bound. In this case, 
\begin{equation}
\CDc = \Cc \times \Dc
\end{equation} 
and the outer bound in \eqref{outer} coincides with $\CDc$.

\item When $P_S(0) (1+ P_S(0)) \geq P_S(1)>P_S(0)$, the smallest achievable distortion in \eqref{ex:distortin} (obtained for $\beta=1$) is $D_{\min}=P_S(1)P_S(0)$. Moreover,  the region $\CDc$ includes 
	 the set of all quadruples $(R_1,R_2,D_1,D_2)$ that for some $\beta,\gamma\in[0,1]$  satisfy	 the rate-constraints 
	\begin{subequations}\label{ex:inner}
		\begin{IEEEeqnarray}{rCl}
			R_k&\leq& 1, \qquad k\in\{1,2\},\\
			R_1+R_2&\leq &	1+ \gamma P_{S}(1) \bigg(H_\textnormal{b}\left(1-\frac{1-\beta}{\gamma}\right)-P_S(0)\bigg), \nonumber \\
		\end{IEEEeqnarray}
	\end{subequations}
and	the distortion constraints in \eqref{eq:Dh}, which simplify to
	\begin{equation}
	D_k \geq \frac{1}{2} (1-\beta)P_S(1) + \beta P_S(1)P_S(0), \quad k=1,2.
	\end{equation}
	
\item When $P_S(1) > P_S(0) (1+ P_S(0))$, the smallest achievable distortion in \eqref{ex:distortin} (obtained for $\beta=0$) is $D_{\min}=\frac{1}{2} P_S(0) (1+P_S(0))$. Moreover, the region $\CDc$ includes 
	 the set of all quadruples $(R_1,R_2,D_1,D_2)$ that for some $\gamma\in[0,1]$  and $\beta \in [0,\gamma]$ satisfy	 the rate-constraints 
	\begin{subequations}	\label{eq:inner2}
		\begin{IEEEeqnarray}{rCl}
			R_k&\leq& 1, \qquad k\in\{1,2\},\\
			R_1+R_2&\leq &	1+ \gamma P_{S}(1) \bigg(H_\textnormal{b}\left(\frac{\beta}{\gamma}\right)-P_S(0)\bigg), \label{eq:lower_sum}
			 \IEEEeqnarraynumspace 
		\end{IEEEeqnarray}
	\end{subequations}
	and the distortion constraints in \eqref{eq:Dh}, which simplify to
	\begin{equation}
	D_k \geq \frac{1}{2} (1-\beta)P_S(0) (1+P_S(0)) + \beta P_S(1)P_S(0), \quad k=1,2
	\end{equation}
	\end{itemize}
		\end{corollary}
	\begin{IEEEproof} Based on Theorem~\ref{outer1} and Proposition~\ref{prp:inner}. See Appendix~\ref{app:B} for details. 
				\end{IEEEproof}
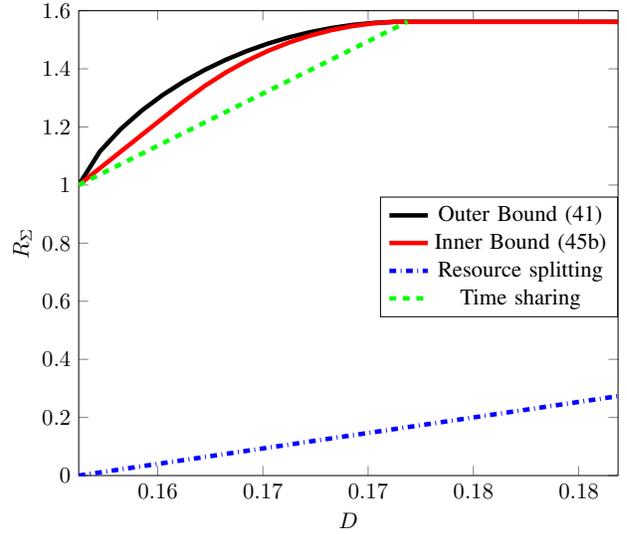
\begin{figure}[!h]
	\centering
	\hspace{-1cm}
	\begin{tikzpicture}[every pin/.style={fill=white},scale=.85]
	\begin{axis}[%
	xlabel={ {$D$ }},
	ylabel={ {$R_{\Sigma}$ }},
	xlabel style={yshift=0.2em},
	ylabel style={yshift=-1em},
at={(1.011in,0.642in)},
scale only axis,
xmin=0.15625,
xmax=0.181875,
ymin=0,
ymax=1.6,
axis background/.style={fill=white},
legend style={at={(1,0.6)},anchor=north east}
]
\addplot [color=black, line width=2.0pt]
  table[row sep=crcr]{%
0.15625	1\\
0.15725	1.11493262742641\\
0.15825	1.19300663526176\\
0.15925	1.25660468269536\\
0.16025	1.31045924881934\\
0.16125	1.35679912448532\\
0.16225	1.39692059113385\\
0.16325	1.43166498432848\\
0.16425	1.46161569767371\\
0.16525	1.48719457078673\\
0.16625	1.50871456997003\\
0.16725	1.52641087358079\\
0.16825	1.54046022361587\\
0.16925	1.55099339848532\\
0.17025	1.55810337593429\\
0.17125	1.5618506139973\\
0.171875	1.5618506139973\\
0.181875	1.5618506139973\\
};
\addlegendentry{Outer Bound \eqref{outer}}

\addplot [color=red,, line width=2.0pt]
  table[row sep=crcr]{%
0.15625	1\\
0.16125	1.28823216598042\\
0.16225	1.34172745484513\\
0.16325	1.38805331243798\\
0.16425	1.42798759689828\\
0.16525	1.46209276104897\\
0.16625	1.49078609329337\\
0.16725	1.51438116477439\\
0.16825	1.53311363148783\\
0.16925	1.5471578646471\\
0.17025	1.55663783457905\\
0.17125	1.5616341519964\\
0.171875	1.5616341519964\\
0.181875	1.5616341519964\\
};
\addlegendentry{Inner Bound \eqref{eq:lower_sum}}

\addplot [color=blue, dashdotted, line width=2.0pt]
  table[row sep=crcr]{%
0.15625	0\\
0.25	1\\
};
\addlegendentry{Resource splitting }

\addplot [color=green, dashed, line width=2.0pt]
  table[row sep=crcr]{%
0.15625	1\\
0.171875	1.5625\\
};
\addlegendentry{Time sharing}

\end{axis}
\end{tikzpicture}%
\vspace{-0.2cm}
\caption{Upper and lower bounds  of Corollary~\ref{ex:general} on the maximum achievable sum-rate  $R_{\Sigma}$ in function of the admissible distortion $D_1=D_2=D$  for the state-dependent Dueck BC when $P_{S}(1)=3/4$ and $P_S(0)=1/4$.}
\label{general_example}
\vspace{-.15cm}

\end{figure}

We  evaluate the bounds for the state distribution  $P_{S}(1)=\frac{3}{4}$ and $P_S(0)=\frac{1}{4}$, which satisfies the condition $P_S(1) \geq P_S(0)(1+P_S(0))$.   
  Specifically, we analyze the largest sum-rates 
	$R_{\Sigma}(D):=R_1+R_2$ that our inner and outer bounds admit under  given symmetric distortion constraints $D_1=D_2 =D$, and 
  compare them to the  baseline schemes. 
	Notice first that for $P_{S}(1)=\frac{3}{4}$ and $P_S(0)=\frac{1}{4}$ the distortion constraint \eqref{ex:distortin} specializes to
	\begin{equation}
	D\geq \frac{1}{2}\left[(1-\beta) \frac{5}{16}+\beta \frac{6}{16}\right]=\frac{5+\beta}{32}, 
	\end{equation}
	and so the minimum distortion (obtained for $\beta=0$)
	
	 is $D_{\min}=\frac{5}{32}$. For $\beta=1/2$ we obtain $D \geq \frac{11}{64}$.  
	Turning back to the  sum-rate $R_{\Sigma}$, for above state distribution, the outer bound \eqref{outer} implies  \begin{equation}
R_{\Sigma}(D) \leq \left\{ \,
\begin{IEEEeqnarraybox}[\IEEEeqnarraystrutmode 
\IEEEeqnarraystrutsizeadd{7pt}
{7pt}][c]{rll}
& 1+\left(\frac{3}{4}\right)^2 H_{\textnormal{b}}(32\cdot D-5), \; \; & \textnormal{if }\frac{5}{32}\leq D\leq \frac{11}{64},
\\
 &\frac{25}{16},\qquad &\textnormal{if }D\geq \frac{11}{64}
\end{IEEEeqnarraybox}
\right.
\label{eq:example_left_right2}
\end{equation}
and the inner bound \eqref{eq:inner2} implies 
\begin{IEEEeqnarray}{rCl}
\lefteqn{R_{\Sigma}(D) \geq }\quad \nonumber \\ 
&&\left\{ \,
\begin{IEEEeqnarraybox}[\IEEEeqnarraystrutmode 
\IEEEeqnarraystrutsizeadd{7pt}
{7pt}][c]{rll}
& 1+ \max_{32D -5 \leq \gamma \leq 1} \frac{3\gamma}{4}\left( H_{\textnormal{b}}\left(\frac{32 D-5}{\gamma}\right)-\frac{1}{4}\right) ,\; \; \\
& &\hspace{-20mm}  \textnormal{if }\frac{5}{32}\leq D\leq \frac{11}{64},
\\
&\frac{25}{16},  &\hspace{-12mm} \textnormal{if }D\geq \frac{11}{64}.
\end{IEEEeqnarraybox}\IEEEeqnarraynumspace
\right.
\label{eq:example_left_right2}
\end{IEEEeqnarray}

Fig.~\ref{general_example} compares these two bounds to the maximum admissible sum-rates $R_{\Sigma}$ attained  by the resource splitting baseline scheme, and by time-sharing the two points of our lower bound \eqref{eq:example_left_right2} that have minimum distortion $(R_\Sigma=1, D=D_{\textnormal{min}}=5/32)$ and maximum rate $(R_{\Sigma}= 25/16, D=11/64)$. 

The resource splitting scheme achieves  $(R_{\Sigma}=0,  D= D_{\min}=5/32)$ during the sensing mode, by setting $X_1= X_2$ (either $0$ or $1$) and not using input $X_0$ at all. (This input  is useless for state sensing.) Moreover, it achieves $(R_\Sigma=1, D=1/4)$ in the communication mode, by completely ignoring the feedback,  sending uncoded bits using inputs $X_0$, and estimating $\hat{S}_1=\hat{S}_2=1$. (This estimator is optimal without feedback  because $P_S(1)> P_S(0)$.)

	\section{Conclusion}

Motivated by a joint radar and communication system, we studied joint sensing and communication over memoryless state-dependent broadcast channels (BC). 
First, we presented a sufficient condition under which there is no tradeoff between sensing and communication. Then, we characterized the capacity-distortion tradeoff region of the physically degraded BC. We further presented inner and outer bounds on the capacity-distortion region of general BCs with states and showed at hand of an example that they  can be tight. Our numerical examples demonstrate that the proposed co-design schemes significantly outperforms the traditional co-exist scheme where resources are split between communication and state sensing. 
	\section*{Acknowledgement}
	M. Ahmadipour and M. Wigger acknowledge
	funding from the ERC under grant agreement 715111. The work of M. Kobayashi is supported by DFG Grant KR 3517/11-1.
	\bibliographystyle{IEEEbib}
	\bibliography{main}	
	

\appendices
\section{Proof of Theorem~\ref{outer1}}\label{app:converse_proof}
	Fix a  sequence of $(2^{nR_1}, 2^{nR_2}, n)$ codes satisfying \eqref{eq:asymptotics}. 
			Fix a blocklenth $n$ and start with Fano's inequality:
			\begin{IEEEeqnarray}{rCl}
				R_1&=&\frac{1}{n} H(W_1)\nonumber\\
				&\leq  & \frac{1}{n}\sum_{i=1}^{n}I(W_1;Y_{1i}, S_{1i}\mid Y_1^{i-1}, S_{1}^{i-1})+\epsilon_n\nonumber\\
				&\leq& \frac{1}{n}\sum_{i=1}^{n}I(W_1, Y_1^{i-1}, S_1^{i-1};Y_{1,i}, S_{1,i} )+\epsilon_n\nonumber\\
				&=&I(W_1, Y_1^{T-1}, S_1^{T-1}; S_{1,T}, Y_{1,T}\mid T)+\epsilon_n\nonumber\\
				&\leq &I(W_1, Y_1^{T-1}, S_1^{T-1},T; S_{1,T}, Y_{1,T})+\epsilon_n\nonumber \\
				& = & I(U;Y_1| S_1) +\epsilon_n\label{eq:R1_bound}
			\end{IEEEeqnarray}
			where $T$  is chosen  uniformly over $\{1,\cdots, n\}$ and independent of $X^n, Y^n_1, Y^n_2,  W_1,W_2,S^n_1, S^n_2$; $\epsilon_n$ is a function that tends to 0 as $n\to \infty$; $U \triangleq (W_1, Y_1^{T-1}, S_1^{T-1},T)$; and $Y_1 \triangleq Y_{1,T}$ and $S_1 \triangleq S_{1,T}$. Notice that $S_1 \sim P_{S_1}$ and it is independent of $(U,X)$, where we define $X\triangleq X_T$. 
			
			Following similar steps, we obtain:
			\begin{IEEEeqnarray}{rCl}
				R_2&=&\frac{1}{n}H(W_2)
				\nonumber\\
				&\leq&\frac{1}{n}I(W_2;Y_2^n, S^n_2)+\epsilon_n
				\nonumber\\
				&\stackrel{(a)}\leq&\frac{1}{n}I(W_2;Y_1^n,  S^n_1, Y_2^n,  S^n_2 
				\mid W_1)+\epsilon_n
				\nonumber\\
				&=&\frac{1}{n}\sum_{i=1}^{n}I(W_2;Y_{1i}, Y_{2i},S_{1i}, S_{2i}\mid Y_1^{i-1},Y_2^{i-1} ,
				\nonumber\\
				&&\hspace{4.4cm}S^{i-1}_1, S^{i-1}_2,W_1)+\epsilon_n\nonumber
				\\
				&\leq& \frac{1}{n}\sum_{i=1}^{n}I(X_i,W_2, Y_2^{i-1} , S_{2}^{i-1}; Y_{1,i}, Y_{2,i}, S_{1,i}, S_{2,i}
				\nonumber \\
				&&\hspace{4.2cm}\mid Y_1^{i-1}, S_1^{i-1}, W_1)+\epsilon_n\nonumber
				\\
				&	=&\frac{1}{n}\sum_{i=1}^{n} I(X_i ; Y_{1,i}, Y_{2,i}, S_{1,i}, S_{2,i} \mid Y_1^{i-1},  S_1^{i-1}, W_1)+\epsilon_n\nonumber
				\\
				&=&I(X_T; Y_{1T}, Y_{2T}, S_{1,T}, S_{2,T} \mid Y_1^{T-1},\nonumber\\
				&&\hspace{4cm} S_1^{T-1}, W_1, T)+\epsilon_n \nonumber \\
				& =& I(X; Y_1,Y_2 \mid S_1,S_2, U) +\epsilon_n, \label{eq:R2_bound}
			\end{IEEEeqnarray}
			where $(a)$ follows by the physically degradedness of the SDMBC and where we defined $Y_2 \triangleq Y_{2,T}$ and $S_{2}\triangleq S_{2,T}$.
			
				Recall that we assume the optimal estimators \eqref{eq:estimator}  in Lemma~\ref{lemma:Shat}. Using the definitions of $T$, $X$, $S_k$ above and  defining $Z\triangleq Z_T$, we can write the average expected distortions as:
			\begin{IEEEeqnarray}{rCl}\label{eq:D_bound}
				\frac{1}{n} \sum_{i=1}^n \EE[d_k(S_{k,i}, \hat{s}_{k,i}^*(X_i,Z_i)] =\EE[ d_k ( S_{k}, \hat{s}_{k,T}^*(X,Z) ]. \IEEEeqnarraynumspace
			\end{IEEEeqnarray}
			
			Combining  \eqref{eq:R1_bound}, \eqref{eq:R2_bound}, and \eqref{eq:D_bound} and letting $n\to \infty$, we obtain that there exists a limiting pmf $P_{UX}$ such that the tuple $(U,X,S_1,S_2,Y_1,Y_2,Z) \sim P_{UX} P_{S_1S_2} P_{Y_1Y_2Z|S_1S_2X}$ satisfies the rate-constraints 
			\begin{IEEEeqnarray}{rCl}
				R_1 & \leq & I(U; Y_1\mid S_1)\\
				R_2 &\leq & I(X; Y_1,Y_2 \mid S_1,S_2, U) 
			\end{IEEEeqnarray}
			and the distortion constraints 
			\begin{IEEEeqnarray}{rCl} 
				\EE[ d_k ( S_{k}, \hat{S}_{k,T}(X,Z) ] \leq D_k, \quad k=1,2,
			\end{IEEEeqnarray}
			for a \emph{possibly probabilistic} estimator $\hat{S}_{k,T}(X,Z)$. Similar to the proof of Lemma~\ref{lemma:Shat} one can however show  optimality of the estimator in \eqref{eq:est}. This complete the  proof.
			
\section{Proof of Corollary~\ref{ex:general}}\label{app:B}
			
			\subsection{Proof of the Outer Bound}		
			The outer bound is based on Theorem~\ref{outer1}, as detailed out in the following. 
		From \eqref{R1} and \eqref{R2} we obtain:
		\begin{IEEEeqnarray}{rCl}
		R_1&\leq& I(U_1; Y_1', X_0\mid S_1, S_2)\nonumber
		\\
		&=&H(U_1)-H(U_1\mid Y_1',X_0,S_1,S_2)
		\nonumber\\
		&=&H(U_1)-H(U_1\mid X_0)
		\nonumber\\
		&=&I(U_1;X_0)		\nonumber\\
		&=& H(X_0)-H(X_0\mid U_1)\nonumber\\
		&\leq& 1-p	,	
	\end{IEEEeqnarray}
 where we defined $p:=H(X_0\mid U_1)$, and 	
	\begin{IEEEeqnarray}{rCl}
		R_2&\leq& I(X_0,X_1,X_2; Y_1', Y_2'\mid S_1, S_2,U_1)
		\nonumber\\
		&\leq&H(X_0\mid U_1)+I(X_1,X_2;Y_1', Y_2'\mid S_1, S_2,U_1)
		\nonumber\\
		&=&H(X_0\mid U_1)+I(X_1,X_2;Y_2'\mid S_1, S_2,U_1)\nonumber
		\nonumber\\
		&&\hspace{2.1cm}+I(X_1,X_2;Y_1'\mid  S_1, S_2,Y_2',U_1)
		\nonumber\\
		&\leq&H(X_0\mid U_1)+P_{S_1S_2}(1,1)\cdot H(X_1 \oplus X_2)
		\nonumber\\
		&=&p+P_{S_1S_2}(1,1) \cdot H_b(\beta).
	\end{IEEEeqnarray}
	where we defined  $\beta := \Pr[X_1 \neq X_2]$.

	In a similar way, we obtain \eqref {eq:R1-outer} and \eqref{eq:R2-outer} from \eqref{upper3} and \eqref{upper4}. 
	
Distortion constraint \eqref{eq:Dh} can be shown by evaluating the optimal estimator in \eqref{eq:est} for this example, as we detail out in the following.
	
	\newcommand{\bx}{\overline{x}}
We first derive the optimal estimator $\hat{s}^{*}_k ((x_1, x_2), z)$ for a given realization of channel inputs and the feedback defined in \eqref{eq:estimator}. Denote the  distortion resulting from this optimal estimator for a given triple $(x_1, x_2, z)$ by
\begin{IEEEeqnarray}{rCl}
d'_k ((x_1, x_2), z)  & =& P_{S_k|X_1X_2Z}( \hat{s}^{*}_k \big ((x_1, x_2), z) \oplus 1 \big |x_1, x_2, z \big)\\
& =& \min_{s} \sum_{s_k\in \{0, 1\}}  (s_k \oplus s)P_{S_k|X_1X_2Z}(s_k|x_1, x_2, z).\nonumber\\
\end{IEEEeqnarray}
The expected distortion can then be expressed as:
\begin{equation}\label{eq:ExpDistortion}
\sum_{x_1, x_2, z} P_{X_1 X_2Z}(x_1, x_2,z)  d'_k((x_1, x_2), z).
\end{equation}
In the following we identify $\hat{s}^*_k ((x_1, x_2), z)$.\\
{\bf Case $z=(1,1)$: } In this case, $S_1=A_2=1$ and 
\begin{align}\label{eq:est-11}
\hat{s}^*_k ((x_1, x_2),( 1,1)) = 1,\;\quad \forall x_1= x_2, \;  k=1,2,
\end{align}
which yields   for any $k=1,2$ and $(x_1,x_2)$:
\begin{equation}\label{eq:dist11}
d'_k((x_1, x_2), (1,1))=0.
\end{equation}
{\bf Case $z=(1,0)$: } In this case, $S_1=1$  and  the optimal estimator produces $\hat{s}^*_1 ((x_1, x_2), (1,0)) = 1$, irrespective of $x_1,x_2$. Consequently, as before, for any  $(x_1, x_2)$:
\begin{equation}\label{eq:dist10_1}
d'_1(x_1, x_2, (1,0))=0. 
\end{equation}
For receiver 2, we distinguish whether  $x_1=x_2$ or $x_1 \neq x_2$.  When, $x_1=x_2$, then $S_2=y_2'=0$ because in this case $x_2\oplus N =x_1 \oplus N$ and this latter equals $1$ because $y'_1=1$. The optimal estimator thus sets $\hat{s}^*_2((x_1, x_2), z)=0$ when $x_1=x_2$, which achieves $0$ distortion $d_2'((x_1, x_2),z)=0$ .

When $x_1\neq x_2$, then  $x_2\oplus N =1 \oplus( x_1 \oplus N)=0$ and the feedback $z$ is independent of the state $S_2$ because this latter is independent of state $S_1$.  The optimal estimator for $x_1\neq x_2$ is thus $\hat{s}^*_2 ((x_1, x_2), (1, 0))=  \onev\{ P_{S} (1) \geq P_{S} (0)\}$.   
This yields the    distortion for any $(x_1,x_2)$:
\begin{align}\label{eq:dist10}
d'_2(x_1, x_2, (1, 0))= \min\{P_{S}(0 ) , P_{S}(1) \} \cdot \onev\{ x_1\neq x_2 \} .
\end{align}	

{\bf Case $z=(0,1)$: } This case is similar to the case $z=(1,0)$ but with exchanged roles for indices $1$ and $2$. So, 
\begin{subequations}\label{eq:dist01}
\begin{align}
d'_1((x_1, x_2), (0,1)) &= \min\{P_{S}(0 ) , P_{S}(1) \} \cdot \onev\{ x_1\neq x_2 \}\\
d'_2((x_1,x_2), (0,1)) &= 0.
\end{align}
\end{subequations}

{\bf Case $z=(0,0)$: }  We again distinguish the two cases $x_1=x_2$ and $x_1\neq x_2$ and start by considering $x_1=x_2$.  In this case,    $x_1\oplus N = x_2 \oplus N$, and so if $S_k=1$ then $Z=(0,0)$ only if $x_{1}\oplus N=x_2 \oplus N=0$, which happens with probability $1/2$. By the independence of the states and the inputs we then have:
\begin{eqnarray*}
\lefteqn{P_{S_k|X_1X_2Z}(1| x_1,x_2, (0, 0))  } \quad \\
& = & \frac{P_{S_k}(1) P_{Y_1'Y_2'|X_1X_2S_k}(0,0|x_1,x_2,1)}{P_{Y_1'Y_2'|X_1X_2}(0,0|x_1,x_2)}\\ 
& = & \frac{P_S(1) 1/2}{P_{Y_1'Y_2'|X_1X_2}(0,0|x_1,x_2)}.
\end{eqnarray*} 
If $S_k=0$, then $z=(0,0)$ happens when $x_1\oplus N=x_2 \oplus N =0$ or when $x_1\oplus N=x_2 \oplus N =1$ and $S_{\bar{k}}=0$, where $S_{\bar{k}}=1-S_{k}$. Since these are exclusive events and have total probability of $1/2 + 1/2\cdot P_S(0)$,  we obtain:
\begin{eqnarray}
\lefteqn{P_{S_k|X_1X_2Z}(0| x_1,x_2, (0, 0))  } \nonumber \quad \\
& = & \frac{P_{S_k}(0) P_{Y_1'Y_2'|X_1X_2S_k}(0,0|x_1,x_2,0)}{P_{Y_1'Y_2'|X_1X_2}(0,0|x_1,x_2)}\\ 
& = & \frac{P_S(0) (1/2+ 1/2 \cdot P_S(0))}{P_{Y_1'Y_2'|X_1X_2}(0,0|x_1,x_2)}.
\end{eqnarray} 
We conclude that for $z=(0,0)$ and $x_1=x_2$, the optimal estimator is 
\begin{equation}
\hat{s}^*_k ((x_1, x_2), (0, 0))   =
\onev\left\{P_{S}(0)(1+P_{S}(0))<P_{S}(1)\right\}, 
\end{equation}
and the corresponding distortion
\begin{IEEEeqnarray}{rCl}\label{eq:dist00equal}
d'_k((x_1, x_2), (0, 0))&=&
 \frac{1}{2}\cdot \frac{\min\left\{ P_S(0) (1+P_S(0)), P_S(1) \right\}}{P_{Y_1'Y_2'|X_1X_2}(0,0|x_1,x_2)} , \nonumber \\
& &  \hspace{3.5cm} \quad  x_1=x_2.\IEEEeqnarraynumspace
\end{IEEEeqnarray}

We turn to the case $x_1\neq x_2$, where $x_1 \oplus N =1 \oplus ( x_2  \oplus N)$. As before, if $S_k =1$, then $Y_k'=0$ only if $x_1\oplus N=0$, which happens with probability $1/2$. Now this implies  $x_2 \oplus N=1$, and thus  $Y_{\bar{k}}'=0$ only if $S_{\bar{k}}=0$, which happens with probability $P_S(0)$. We thus obtain  for $x_1 \neq x_2$:
\begin{eqnarray}
\lefteqn{P_{S_k|X_1X_2Z}(1| x_1,x_2,( 0, 0))  } \quad \nonumber \\
& = & \frac{P_{S_k}(1) P_{Y_1'Y_2'|X_1X_2S_k}(0,0|x_1,x_2,1)}{P_{Y_1'Y_2'|X_1X_2}(0,0|x_1,x_2)}\\ 
& = & \frac{P_S(1) 1/2 P_{S}(0)}{P_{Y_1'Y_2'|X_1X_2}(0,0|x_1,x_2)}.
\end{eqnarray} 
If $S_k=0$, then $z=(0,0)$ happens when $x_{\bar{k}} \oplus N =0$ or when $x_{\bar{k}} \oplus N =1$ and $S_{\bar{k}}=0$. Since these are exclusive events with total probability $1/2+1/2 \cdot P_S(0)$, we obtain:
\begin{eqnarray}
\lefteqn{P_{S_k|X_1X_2Y_1'Y_2'}(0| x_1,x_2, 0, 0)  } \quad \nonumber \\
& = & \frac{P_{S_k}(0) P_{Y_1'Y_2'|X_1X_2S_k}(0,0|x_1,x_2,0)}{P_{Y_1'Y_2'|X_1X_2}(0,0|x_1,x_2)}\\ 
& = & \frac{P_S(0) (1/2+ 1/2 \cdot P_S(0))}{P_{Y_1'Y_2'|X_1X_2}(0,0|x_1,x_2)}.
\end{eqnarray} 
We conclude that for $z=(0,0)$ and $x_1\neq x_2$, the optimal estimator is
\begin{align}\label{eq:OptEstimator00}
\hat{s}^*_k ((x_1, x_2), (0, 0))  
= \onev\{(1+ P_{S}(0)) <  P_{S}(1) \},   \quad  x_1\neq x_2, 
\end{align} 
and the corresponding distortion
\begin{align}\label{eq:dist00}
 d'_k((x_1, x_2), (0, 0))
&=
 \frac{1}{2}\cdot \frac{P_S(0)\min\{(1+ P_{S}(0)), P_{S}(1) \} }{P_{Y_1'Y_2'|X_1X_2}(0,0|x_1,x_2)}, \nonumber \\
 & \hspace{4cm}    x_1\neq x_2.
\end{align}

We now turn to the conditional probabilities of the feedbacks given the inputs that are required to evaluate \eqref{eq:ExpDistortion}. Whenever the inputs $x_1=x_2$, 
\begin{align}\label{eq:1}
P_{Y_1'Y_2'|X_1 X_2} (0,0|x_1,x_2) &=\frac{ 1+(P_{S}(0))^2 }{2}, 
\end{align}
because $Y_1'=Y_2'=0$ happens only when either $N= x_1   = x_2$ or when $N=x_1\oplus 1=x_2\oplus 1$ and $S_1=S_2=0$. These two are exclusive events and happen with total probability $1/2 + 1/2 (P_S(0))^2$.  
Whenever,  $x_1\neq x_2$: 
\begin{align}
P_{Y_1'Y_2'|X_1X_2} (0,1|x_1, x_2) &=
P_{Y_1' Y_2'|X_1X_2} (1,0|x_1, x_2) \\
&=  \frac{P_{S}(1)}{2}, 
\end{align}
by symmetry and because  for $x_1\neq x_2$ the event $Y_1'=1$ and $Y_2=0$ happens only when $S_1=1$ and $N= x_1\oplus 1$. (Notice that since $x_1\neq x_2$, this latter condition implies that $N \oplus x_2=0$ and thus $Y_2'=0$ independent of $S_2$.)
Moreover, when $x_1\neq x_2$: 
\begin{align}\label{eq:3}
P_{Y_1', Y_2'|X_1, X_2} (0,0|x_1, x_2) &= P_{S}(0),
\end{align}
because  for $x_1 \neq x_2$, the event $Y_1'=0$ and $Y_2'=0$ happens when either $N = x_1= x_2 \oplus 1$ and $S_2=0$ or when $N=x_1 \oplus 1=x_2$ and $S_1=0$.  These are exclusive events and happen with total probability $1/2 P_S(0)+1/2 P_S(0)=P_S(0)$. 
	
	Plugging \eqref{eq:dist10_1}, \eqref{eq:dist10}, \eqref{eq:dist01}, \eqref{eq:dist00equal}, \eqref{eq:dist00}  and \eqref{eq:1}--\eqref{eq:3} into \eqref{eq:ExpDistortion} establishes the desired distortion constraint \eqref{eq:Dh} and  concludes the proof of the outer bound.

\subsection{Proof of Achievability Results}
		
The achievability results can be obtained by evaluating  Proposition~\ref{prp:inner} for the following choices:  $X_0, X_1, X_2$  Bernoulli-$\frac{1}{2}$ with $X_0$ independent of $(X_1,X_2)$ and $X_1=X_2=x$ with probability  $\frac{1-\beta'}{2}$ for all $x\in\{0,1\}$; $U_i=X_i$,  for $i=0, 1, 2$; and one of the following three choices: $V_1=(X_0,X_1)$, $V_2=(X_0,X_2)$,    $V_0=X_1\oplus Y_1'$ or   $V_1=(X_0,X_1)$, $V_2=(X_0,X_2)$, $V_0=X_2\oplus Y_2'$ or $V_1=V_2=V_0=0$. The last choice corresponds to not using  feedback for communication and achieves all quadruples $(R_1,R_2,D_1,D_2)$ satisfying $R_1+R_2 \geq 1$ and $D_k \geq D_{\min}$, where the value of  $D_{\min}$ depends on the state probabilities  $P_S(0)$ and $P_S(1)$ and is specified in the theorem.

More specifically,  achievability of $\Cc\times \Dc$ when $P_S(1)\leq P_S(0)$ can be established by  time-sharing between the first two choices where we set $\beta'=0$ in both of them. (That means we choose $X_1$ and $X_2$ to be independent.)   

Achievability of \eqref{ex:inner} can be established by time-sharing  one of the first two choices with parameter $\beta' = 1-\frac{1-\beta}{\gamma}$ over the fraction $\gamma$ of time with the third choice over the remaining  fraction $1-\gamma$ of time. 

Achievability of \eqref{eq:inner2} can be established by time-sharing  one of the first two choices with parameter $\beta' = \frac{\beta}{\gamma}$ over the fraction $\gamma$ of time with the third choice over the remaining  fraction $1-\gamma$ of time.


\end{document}